\documentclass[a4paper]{jpconf}

\usepackage{graphicx}

\usepackage[english]{babel}
\usepackage{aasjournals}
\usepackage[hidelinks]{hyperref}
\usepackage[sort&compress,numbers]{natbib}

\usepackage{amsfonts}
\usepackage{amsmath}
\usepackage{amssymb}
\usepackage{graphics}
\usepackage{xcolor}
\usepackage{setspace}
\usepackage{verbatim}
\usepackage{listings}
\usepackage{subcaption}

\definecolor{codegreen}{rgb}{0,0.6,0}
\definecolor{codegray}{rgb}{0.5,0.5,0.5}
\definecolor{codepurple}{rgb}{0.58,0,0.82}
\definecolor{backcolour}{rgb}{0.95,0.95,0.92}

\lstdefinestyle{mystyle}{
    backgroundcolor=\color{backcolour},   
    commentstyle=\color{codegreen},
    keywordstyle=\color{magenta},
    numberstyle=\tiny\color{codegray},
    stringstyle=\color{codepurple},
    basicstyle=\ttfamily\footnotesize,
    breakatwhitespace=false,         
    breaklines=true,                 
    captionpos=b,                    
    keepspaces=true,                 
    numbers=left,                    
    numbersep=5pt,                  
    showspaces=false,                
    showstringspaces=false,
    showtabs=false,                  
    tabsize=2
}
\newcommand{\pynucastro}{{\sffamily pynucastro}}
\newcommand{\amrex}{{\sffamily AMReX}}
\newcommand{\reaclib}{{\sffamily REACLIB}}
\newcommand{\castro}{{\sffamily Castro}}
\newcommand{\maestroex}{{\sffamily MAESTROeX}}

\begin{document}
\title{\pynucastro~2.1: an update on the development
of a python library for nuclear astrophysics}

\author{Alexander Smith Clark$^{1}$, Eric T. Johnson$^{1}$, Zhi Chen$^{1}$, Kiran Eiden$^{3}$, Michael Zingale$^{1}$, Brendan Boyd$^{1,2}$, Parker T. Johnson$^{4}$, Luis Rangel DaCosta$^{5}$}

\address{$^1$ Department of Physics and Astronomy, 
Stony Brook University, Stony Brook, NY 11794-3800, USA}
\address{$^2$ Institute for Advanced Computational Science,
Stony Brook University, Stony Brook, NY 11794-5250, USA}
\address{$^{3}$ Department of Astronomy,
University of California, Berkeley, Berkeley, CA 94720-3411, USA}
\address{$^4$ Department of Physics and Astrophysics, 
University of North Dakota, Grand Forks, ND 58202, USA}
\address{$^{5}$ Department of Materials Science and Engineering,
University of California, Berkeley, Berkeley, CA 94720-1760, USA}

\ead{alexander.smithclark@stonybrook.edu}

\begin{abstract}
\pynucastro\ is a python library that provides visualization and
analyze techniques to classify, construct, and
evaluate nuclear
reaction rates and networks.  It provides tools
that allow users to determine the importance of each rate in the
network, based on a specified list of thermodynamic properties.  Additionally, \pynucastro\ can output
a network in C++ or python for use in simulation
codes, include the \amrex-Astrophysics simulation suite. 
We describe the changes in \pynucastro\
since the last major release, including new capabilities
that allow users to generate reduced networks and thermodynamic tables for
conditions in nuclear statistical equilibrium.
\end{abstract}

\section{Introduction}
In stars, nuclear reactions are the primary source of energy, and their
composition drives their evolution \cite{arthur:1920, bethe:1939}.
Understanding stars means understanding how nuclear reactions work, and
their impact on the whole network under different environments and conditions.  Simulations
of stars model their steady evolution and explosive deaths,
requiring accurate and efficient nuclear reaction networks.  In the
current exascale era, GPU-enabled reaction networks are also required.
\pynucastro\ was developed to meet this need---it provides a framework
for exploring reaction rates and networks interactively in python and
for exporting a reaction network to python or (GPU-enabled) C++ code
for use in a simulation.  \pynucastro\ complements other reaction
libraries in the field, for example, {\tt SkyNet} \cite{lippuner:2017}
and {\tt WinNet} \cite{winet:2023}, but its goals differ slightly: the
emphasis in \pynucastro\ is on interactive exploration of a network,
to allow the user to create a network tailored to a particular science
problem, and on the ability to export the network in a format that can
be used in a multidimensional simulation code.  This also allows us to
easily explore the effect of the network size on a simulation, as
done, for example, in a recent study of flames in X-ray
bursts~\cite{Chen_2023}.

Recently \pynucastro\ hit the 2.0 milestone \cite{pynucastro2}, with
the addition of weak rates, a nuclear statistical equilibrium solver,
the ability to derive reverse rates through detailed balance, support
for approximate rates, C++ output for the \amrex-Astrophysics suite of
simulation codes, and some basic methods to help understand the
importance of different rates in a network.  The development has since
continued, with a lot of new features leading to the 2.1 release,
which we summarize here.

\section{Overview of New Features}

\subsection{Expanded weak rates}

The sensitivity of weak rates in core-collapse supernovae, novae, X-ray bursts, and massive stars astrophysical objects and environments plays a significant role in their nucleosynthesis and energy generation \cite{sullivan:2016, fujimoto1981}. In \pynucastro~2.0, we included the weak-rate tabular data \cite{suzuki:2013, suzuki:2016} for nuclei $A=17$ to $28$. This was driven by our desire to
simulate convective Urca \cite{Calder_2019}.  From
these tables, we extracted the
$\beta^-$-decay rate $(\beta^-)$, electron-capture rate ($e^{-}$),
gamma-energy rate ($\Gamma_e$), and the neutrino-energy rate loss
($\Gamma_\nu$), reformatted in cgs units. We know that both weak reactions, $\beta^-$-decay and electron-capture, produce the following changes in a nucleus $(Z,A)$:
\begin{align}
    (Z,A) &\rightarrow (Z+1,A) +e^- +\nu_{e} \nonumber,\\
    (Z,A) + e^- &\rightarrow (Z-1,A) +\bar{\nu}_{e},
\end{align}
However, the positron-capture and the $\beta^+$-decay reactions produce the same effects on the nuclei $(Z,A)$:
\begin{align}
    (Z,A) + e^+ &\rightarrow (Z+1,A) + \bar{\nu}_{e}, \nonumber\\
    (Z,A) &\rightarrow (Z-1,A) + e^+ + \nu_{e}, 
\end{align}
respectively. These additional channels become important at hotter and denser conditions than seen in the Urca simulations, and some rate sources tabulate them in addition to their electron counterparts.  In \pynucastro~2.1, we combine the bare $\beta^-$-decay and the positron-capture rates in a single rate, and similarly, the bare electron-capture and the  $\beta^+$-decay rates.  This means that each sequence $(Z, A) \rightarrow (Z+1,A)$ and $(Z,A) \rightarrow (Z-1,A)$ are represented by a single effective link in a network.

\begin{figure}
\centering
\begin{subfigure}{0.45\textwidth}
    \includegraphics[width=\textwidth]{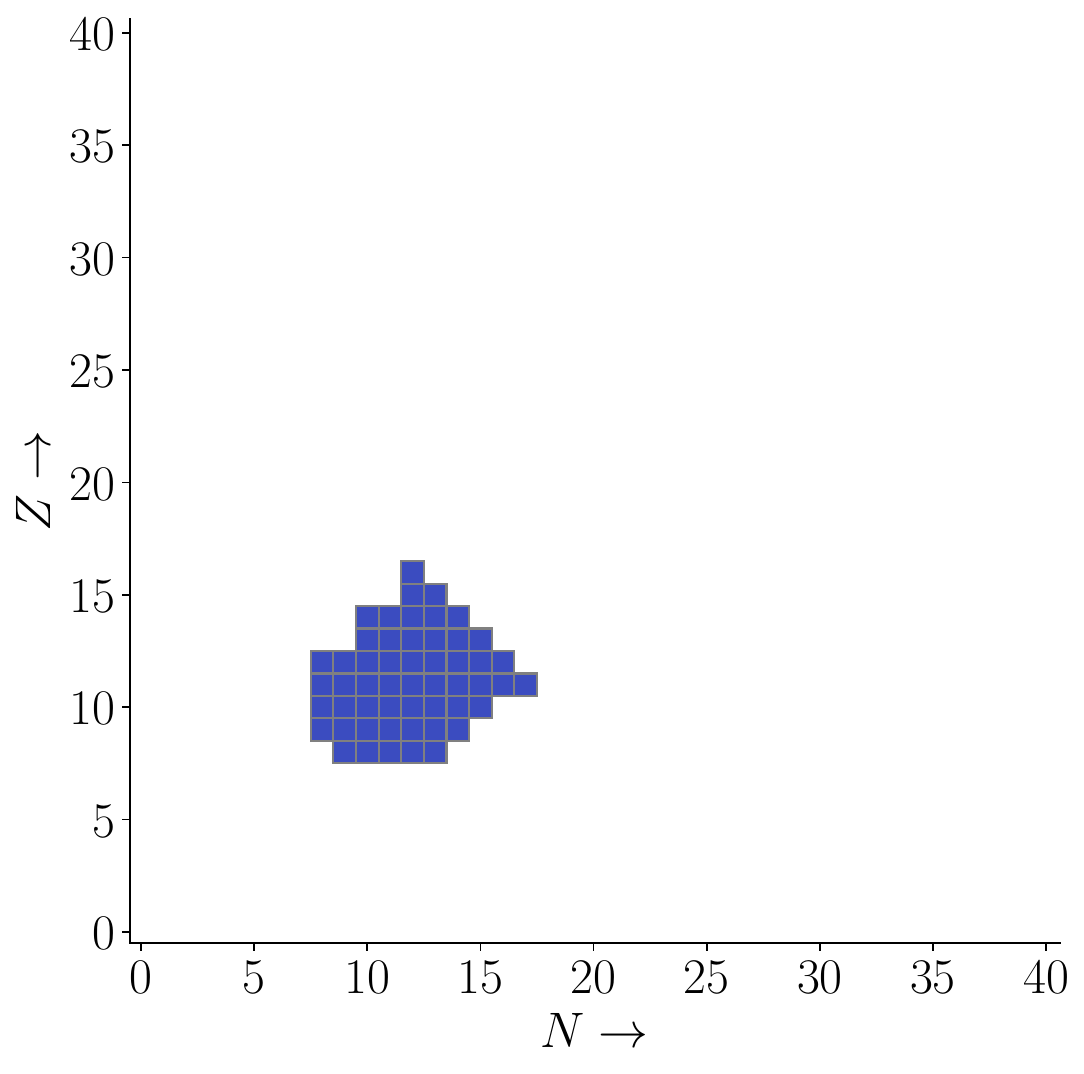}
    \caption{\pynucastro~2.0 weak-rate nuclei availability}
\end{subfigure}
\hspace{0.2cm}
\begin{subfigure}{0.45\textwidth}
    \includegraphics[width=\textwidth]{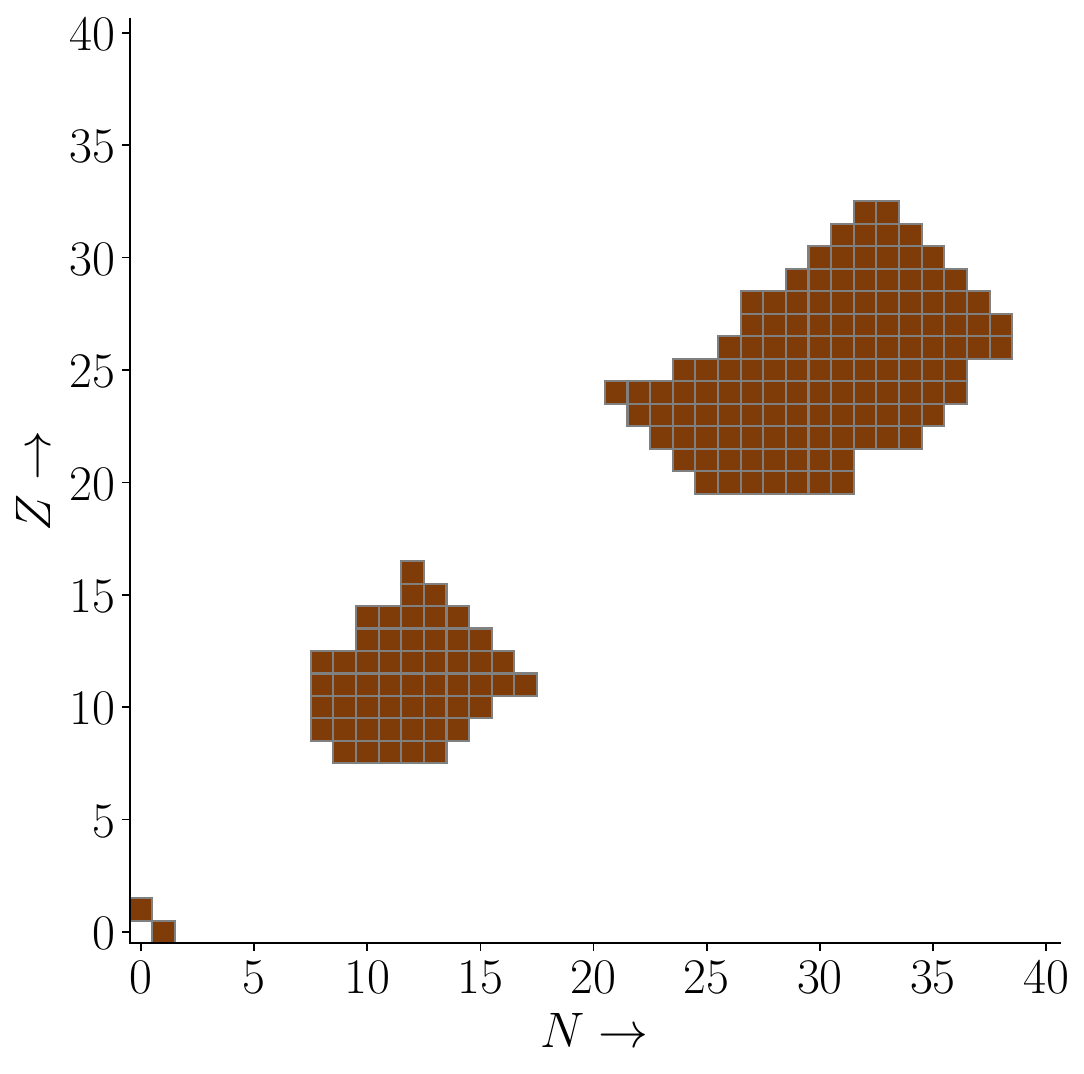}
    \caption{\pynucastro~2.1 weak-rate nuclei availability}
\end{subfigure}
\caption{\small (a) {\it Left}: the original weak rate group of nuclei supported \pynucastro~2.0. (b) {\it Right}: the original weak rate group of nuclei plus the contribution from the \cite{langanke:1999, langanke:2000, langanke:2001} rates. }
\label{fig:1}
\end{figure}

In \pynucastro~2.1, we have improved our weak-rate tabular data availability by including the $A=45$ to $65$ rates in
\cite{langanke:1999, langanke:2000, langanke:2001} to support simulations of massive stars.  We also reformatted our
tables to consider the original logged tabular entries instead, in cgs
units, and in our effective rate convention: $\log(\beta^-)$, $\log(e^{-})$, $\log(\Gamma_e)$, and
$\log(\Gamma_\nu)$ to each tabulated pair $(\log\rho Y_e,\, \log
T)$. A comparison between \pynucastro~2.0 and \pynucastro~2.1 weak rate nuclei availability is depicted in Fig \ref{fig:1}.  Our addition of rates is driven by our science goals, and more rates will be added as needed.

\subsection{A new bilinear interpolation scheme for the tabular rates implementation}

In {\tt pynucastro} 2.0, the tabular rate
evaluation in a {\tt PythonNetwork} simply
returned the closest tabular entry for the $(\rho Y_e, T)$ input pair, while the {\tt AmrexAstroCxxNetwork} C++ network class performed a bi-linear interpolation over the tabular input pairs. Thus, the {\tt PythonNetwork} functionality was made consistent with the more accurate approach in {\tt AmrexAstroCxxNetwork}. 

In our new release, we reformatted our tabular entries in terms of $\log_{10}(\rho Y_e)$ and $\log_{10}(T)$ and implemented a bi-linear interpolation scheme for both network classes. For the {\tt PythonNetwork} class, we implemented a bi-linear interpolation
scheme stored in the {\tt TableInterpolator} class
that can be used both for interactive evaluation
of the rates and when the network is written out to a python module for use elsewhere.  An example
of the new interpolation interface is shown in the following script:

\begin{lstlisting}[language=Python]
def na23__ne23(rate_eval, T, rhoY):
    # na23 --> ne23
    na23__ne23_interpolator = TableInterpolator(*na23__ne23_info)
    r = na23__ne23_interpolator.interpolate(np.log10(rhoY), np.log10(T), 
                                            TableIndex.RATE.value)
    rate_eval.na23__ne23 = 10.0**r
\end{lstlisting}
Note, that we construct first the {\tt TableInterpolator} object for the {\tt na23\_\_ne23} rate, and then we evaluate the rate.
 
\subsection{A Simpler C++ Network}

The original C++ network support in \pynucastro\ is through {\tt AmrexAstroCxxNetwork}, and utilizes
the data structures in the \amrex\ library~\cite{amrex_joss}
to provide the ability to offload the reaction network onto GPUs.  This is compatible with
any simulation code that uses \amrex, like \castro~\cite{castro_joss} or \maestroex~\cite{maestroex-joss}.

\pynucastro\ now supports a pure C++
network, through the {\tt SimpleCxxNetwork} class, that can output the RHS and Jacobian
of the network.  This was developed
to allow other simulation codes to use \pynucastro\ networks.  At the moment, screening is not implemented (but it is easy to add a hook for a user-supplied screening function) and tabulated rates are not supported.  For codes written in C, the C++ interfaces can be accessed via an {\tt extern "C" \{\}} interface.  This network output will be expanded as community demand dictates and we hope this new interface will help spur
the adoption of \pynucastro-generated networks by other
simulation codes.

\subsection{Support for custom rates}

The base {\tt Rate} class in \pynucastro\ was improved to make it easier to add a custom
rate.  This can allow a user to explore variations
or approximations to the standard rates in
the \pynucastro\ library.  For example, 
suppose we want to add a simple powerlaw rate:
\begin{equation}
r = r_0 \left (\frac{T}{T_0} \right )^\nu
\end{equation}
This can be accomplished by creating
a new class that inherits from {\tt Rate},
stores the parameters $r_0$ and $T_0$, and
calls the super-class initialization,
e.g.

\begin{lstlisting}[language=Python]
class MyRate(pyna.Rate):
    def __init__(self, reactants=None, products=None,
                 r0=1.0, T0=1.0, nu=0):

        # we set the chapter to custom so the network knows how to deal with it
        self.chapter = "custom"

        # call the Rate init to do the remaining initialization
        super().__init__(reactants=reactants, products=products)

        self.r0 = r0
        self.T0 = T0
        self.nu = nu
\end{lstlisting}
This sets up all of the logic needed to understand
how to use our new rate, {\tt MyRate}, in a network.
Then
we only need to add a few additional methods.  If we only
want to use the rate in python, then
we just need to write an {\tt eval()} method
that computes the rate given temperature,
such as:
\begin{lstlisting}[language=Python]
    def eval(self, T, rhoY=None):
        return self.r0 * (T / self.T0)**self.nu
\end{lstlisting}
If we want to export the network to python or C++ code,
then we need to add one additional method that
writes the function string.  This custom rate can
then be used in any function in \pynucastro\ that takes
a {\tt Rate} object.

\subsection{The DRGEP modified reduction method implementation}

In \pynucastro~2.0, we provided a simple {\tt find\_unimportant\_rates} method to help reduce the
size of a network.  In the latest release, a more
extensive reduction method framework is included.

The idea of selecting the most important nodes of a graph is also a problem
with important applications in combustion and chemical kinetics
\citep{tianfeng:2005, tianfeng:2006, pepiot:2008}. The general idea
behind these methods starts with a large-sized directed graph
$\mathcal{G}$ representing the network and a user-defined selection of fixed target nodes
$T$. For each pair of nodes $(A,B)$ of the graph $\mathcal{G}$, one can define the following matrix coefficients $r_{A,B}$:
\begin{equation}
    r_{A,B} = \dfrac{\left| \sum_{i=1}^{N_{\mathrm{max}}} \nu_{A,i}\omega_i\delta_{i,B}\right|}{\max(P_A,C_A)}.
\end{equation}
Here, $P_A$ are all the rates on which $A$ is a product, similarly $C_A$ are all the rates on which $A$ is a reactant, $\nu_{A,i}$ is the stoichiometric 
 coefficient of $A$ in the reaction $i$, $\omega_i$ is the $i$-th rate evaluation, and $\delta_{i,B}$ is the Kronecker delta object. From each target nucleus $T$, a weight $R_{TB}$ is defined for each connected nuclei $B$ to the target $T$ as follows:
\begin{equation}\label{weight}
  R_{TB} = \max_{\mathrm{all\; path\; p}}\left(\prod_{i=1}^{n-1} r_{i, i+1, p}\right)  
\end{equation}
where the index $1\leq i \leq n$ represents all the nuclei between $T$ and $B$ for a given path $p$, including both. The weights $R_{TB}$ can be calculated using a modified Dijkstra's algorithm \cite{pepiot:2008, niemeyer:2011, dijstra:1959}. From (\ref{weight}), all of the paths from a nucleus $B$ to a given target $T$ will have weights less than $R_{TB}$. For each target $T$, we can flag whether a nucleus $B$ is considered unimportant by comparing to a user-specified tolerance choice $\epsilon_{T}$:
\begin{equation} \label{check}
    R_{TB} < \epsilon_T.
\end{equation}
We can then remove nuclei with $R_{TB}$ less than $\epsilon_T$ for all possible targets. The reduced network will be a sub-graph $\mathcal{G}'\subset \mathcal{G}$ that will contain at least $T$ nodes:
\begin{equation}
    \mathcal{G} \rightarrow (\mathcal{G}', T)
\end{equation}
Following the particular case of the DRGEP modified method described
in \cite{pepiot:2008, niemeyer:2011, dijstra:1959}, we have implemented the DRGEP
modified reduction method in \pynucastro. A direct application of this method is to identify all of the connections that are important for a particular subset of nuclei from a known reaction network set of nuclei. By setting each nucleus from the known network as a target, we can explore which rates are significant from all the currently supported rate libraries, up to an endpoint nucleus given by the user. For example, let us explore how several nuclei may be connected from the known CNO cycle to all the species up to $^{56}\mathrm{Ni}$. The importance of $^{56}\mathrm{Ni}$ endpoint species is explained, for example in \cite{alexakis:2004}. We start by creating a pool of rates with all the currently supported \reaclib~rates up to $^{56}\mathrm{Ni}$ with a fixed solar composition. This two-step process is achieved by the {\tt load\_network()} method with a fixed network {\tt endpoint} in $^{56}\mathrm{Ni}$.  This will create a {\tt PythonNetwork} containing all the nuclei with $A$ and $Z$ less than the endpoint and all the rates connecting them.  Then, by selecting 25 representative nuclei of the original hydrogen-CNO burning reaction network, we construct the {\tt targets} list. Finally, each link that is ultimately connected to each target, will be evaluated by a weight-constraint function. The {\tt tols} list contains a tolerance that will determine whether each link should be severed or not based on the previous evaluation. From the original pool network composed of a total of 6758 rates with 584 nuclei, we can drastically reduce it to 607 rates with 63 nuclei. Thus, from all the rates \reaclib~ rates with nuclei up to $^{56}\mathrm{Ni}$, we were able to extract the most important rates connected to the original network nuclei targets. 
\begin{lstlisting}[language=Python]
from pynucastro import Nucleus
from pynucastro.reduction import drgep
from pynucastro.reduction.generate_data import dataset
from pynucastro.reduction.load_network import load_network

net = load_network(endpoint=Nucleus('ni56'),
                   library_name='reaclib_default2_20220329')

b_rho = (1e2, 1e6)  # density (g/cm^3)
b_T = (1.0e7, 3.0e8)  # temperature (K)

data = list(dataset(net, n=10, permute=False, b_rho=b_rho, b_T=b_T, b_Z=None))

targets = ["p", "d", "he3", "he4", "li7", "be7", "be8", "b8", "c12",
           "n13", "n14", "n15", "o14", "o15", "o16", "o17", "o18",
           "f17", "f18", "f19", "f20", "ne18", "ne19", "ne20", "ne21"]

targets = [Nucleus(n) for n in targets]

# The tols, for each target, indicate if the nuclei connected to it is
# important or not

tols = [0.3, 0.3, 0.3, 0.3, 0.3, 0.3, 0.3, 0.3, 0.3, 0.3, 0.3, 0.3, 0.3, 
        0.3, 0.3, 0.3, 0.3, 0.3, 0.3, 0.3, 0.3, 0.5, 0.5, 0.5, 0.5]

reduced_net = drgep(net=net, conds=data, targets=targets, tols=tols, 
                    returnobj='net', use_mpi=False, use_numpy=True)
            
\end{lstlisting}

\begin{figure}
\centering
\begin{subfigure}{0.45\textwidth}
    \includegraphics[width=\textwidth]{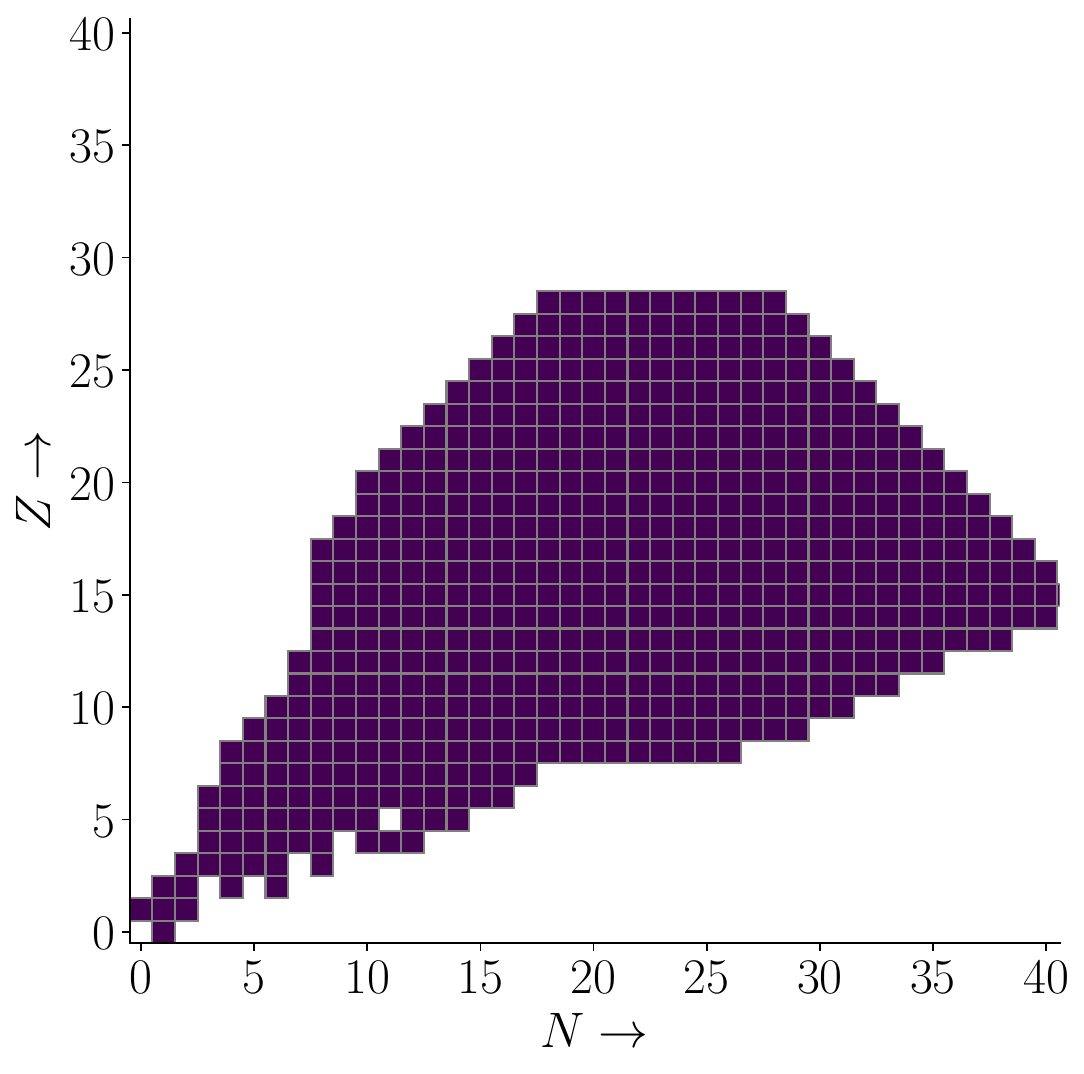}
    \caption{The available pool of rate nuclei}
\end{subfigure}
\hspace{0.2cm}
\begin{subfigure}{0.45\textwidth}
    \includegraphics[width=\textwidth]{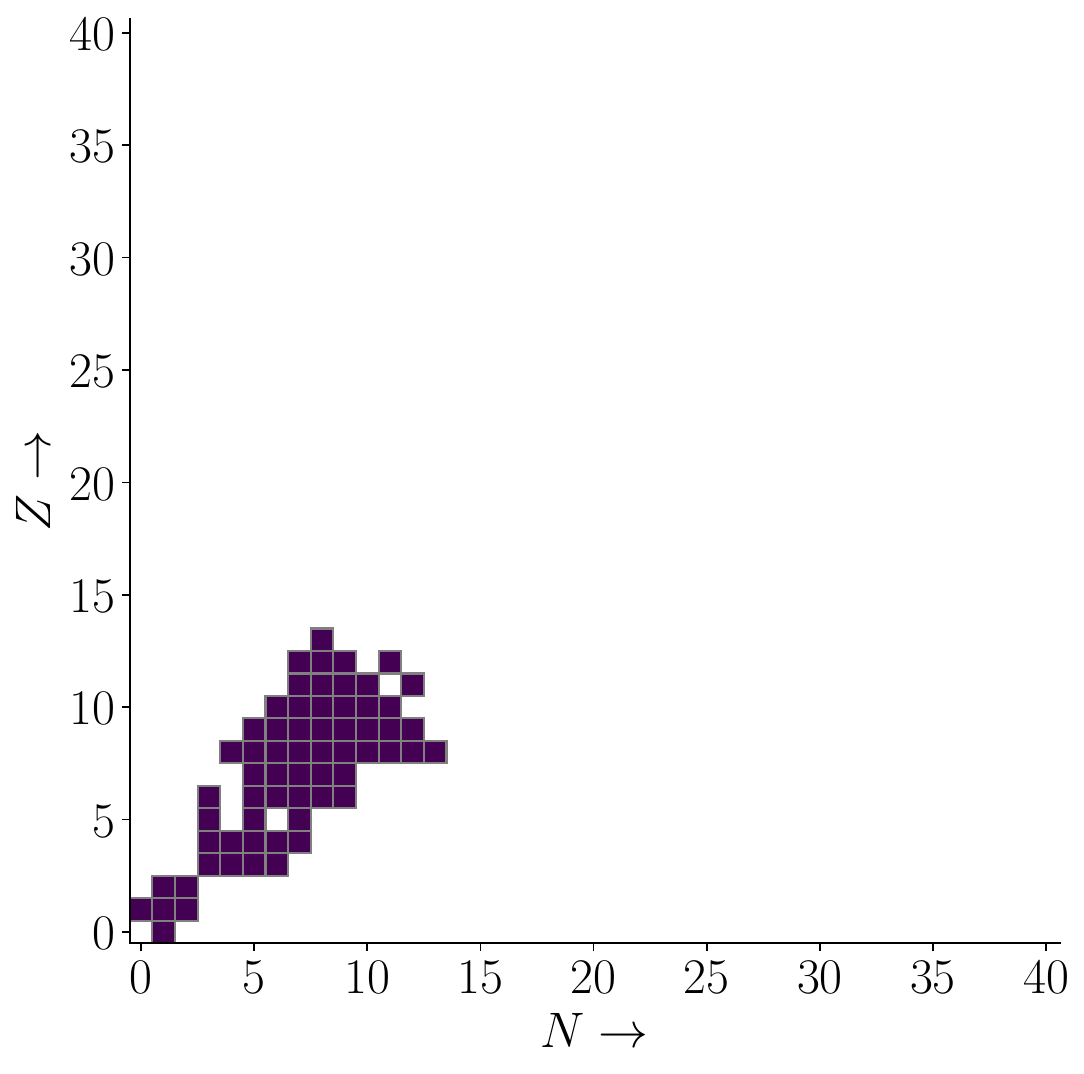}
    \caption{The reduced network nuclei}
\end{subfigure}
\caption{\small (a) {\it Left}: the available rate nuclei up to $^{56}\mathrm{Ni}$, extracted from \reaclib. (b) {\it Right}: the reduced network availability, by selecting a network up to $^{21}\mathrm{Ne}$.}
\label{fig:2}
\end{figure}

An exploration of alternative reduction method implementations, energy-based weight-constraints functions, and the analysis of their implications in large nuclear
reaction networks should be studied further. These algorithms may
prove to be useful in highlighting the most important species and
rates from an extensive nuclear reaction network, improving our
understanding of the group of reactions that dominate our problem
and the efficiency of our calculations.

\subsection{Interfacing with Julia}

The SciPy library \cite{scipy2020} provides an ODE integration function
{\tt solve\_ivp()} that provides many different ODE solvers, including
BDF solvers, tailored for stiff systems.  This is the main way we
suggest how a {\tt PythonNetwork} network should be integrated. However, there are many ODE methods that are not implemented in SciPy, some of which can be more efficient for our problems.

The Julia library {\tt DifferentialEquations.jl} \cite{diffeqjl}
expands the possibilities to consider.  In \pynucastro~2.1, we demonstrated how
to integrate a python network using the Julia solvers. Performance
relies heavily on Numba acceleration \cite{numba}. This opens the
door to further exploration of different ODE solvers. 

This interface makes it straightforward to test the variety of solvers on the network of interest. Below is an example using a simple CNO nuclear reaction network with an initial composition of a standard solar mix. 

\begin{lstlisting}[language=Python]
import numba
import numpy as np
from diffeqpy import de
import cno_test_integrate as cno # importing network created

rho = 150. #g/cm^3
T = 1.5e7  #K
screen_func = None

Y0 = np.zeros((cno.nnuc), dtype=np.float64)
Y0[cno.jp] = 0.7 # Solar mix of mass fractions
Y0[cno.jhe4] = 0.28
Y0[cno.jc12] = 0.02

Y0[:] = Y0[:]/cno.A[:] # Convert to molar fractions

tspan = (0.0, 1.e20)
p = np.array([rho, T]) # initial parameters
alg = de.FBDF() # Choosing Julia ODE solver

def f(u, p, t): 
   """u = array of unknowns, p = array of parameters, t = time"""
    rho, T = p
    Y = u
    return cno.rhs_eq(t, Y, rho, T)

def f_jac(u, p, t):
    rho, T = p
    Y = u
    return cno.jacobian_eq(t, Y, rho, T)

# create Numba compiled versions of the functions
numba_f = numba.njit(f) 
numba_jac = numba.njit(f_jac)
ff = de.ODEFunction(numba_f, jac=numba_jac)

prob = de.ODEProblem(ff, Y0, tspan, p) 
sol = de.solve(prob, alg, abstol=1.e-9, reltol=1.e-9)
\end{lstlisting}
This code integrates the system over the specified time using the {\tt FBDF} solver in the Julia package. Note that the Julia interface requires that the righthand side and Jacobian functions should be combined into a single interface via {\tt ODEFunction}. After integration, {\tt sol.stats} can provide a variety of different statistics including the number of function evaluations iterations, and linear algebra operations it performs. These can be useful in determining the performance of the solver. 

\subsection{NSE table construction}

\pynucastro\ 2.0 saw the addition of a robust NSE solver.
In version 2.1, the NSE solver has been split off into its own network type, {\tt NSENetwork},
to allow for the addition of new features that support creating
NSE tables 
for use in simulation codes.
To support this capability, a new method {\tt Composition.bin\_as()} has been added to bin a large {\tt Composition}, for example, representing the NSE state from 100 nuclei, down to a smaller {\tt Composition} that will be what is carried on the grid in a simulation code.
We've also added the ability to supply an optional rate filter function when evaluating the RHS of a network.  For the NSE evolution, this can be used to only compute the evolution due to weak interactions, which cause an evolution in $Y_e$ further evolving the NSE state.
Likewise, the neutrino loss from weak rates can be returned separately when evaluating the energy production in a network.  Together these new methods allow us
to easily create an NSE table like that in \cite{seitenzhal:2009},
and an interface for this table creation is under development.

\section{Summary and Future Work}
The new release of \pynucastro\ added many new features: an expanded weak rate library with a new interpolation scheme over the tabular rates, a new template to cover bare C++ support, the ability to construct a custom rate, network reduction capabilities, a new Julia integration interface, and the capability of creating an NSE table from the network. All of these new features are driven
by the science applications that the developers are working on, including studies of convective burning in  massive stars leading
up to core-collapse.

The next major developments will likely be the addition of
the remaining pieces needed to support a self-heating burn
completely in python.  This includes the addition of an equation
of state (to get the temperature evolution) and neutrino cooling
sources.  Pull requests with initial implementations of each are
under review, with the equation of state based on 
the Helmholtz equation of state \cite{timmes:1999, timmes_swesty:2000}.
Once the previous step is completed, we may explore the addition of other equations of state, for example \cite{skye:2021}.

Finally, all of the major machine learning libraries have python interfaces and
much AI research relies on these libraries.  By providing a clean,
extensible python interface to nuclear reaction rates and networks, we
enable the use of \pynucastro\ for machine-learning applications.  In
particular, since reaction network integration can be one of the most
time-consuming parts of a simulation, using neural networks to augment
or replace the ODE integration is a promising area \cite{maestro-nn}.

\ack 

The work at Stony Brook was supported by DOE/Office of Nuclear Physics
grant DE-FG02-87ER40317. LRD and KE were supported by the U.S. Department of Energy, Office of Science, Office of Advanced Scientific Computing Research, Department of Energy Computational Science Graduate Fellowship under Award Number DE-SC0021110.  This research has made use of
NASA's Astrophysics Data System Bibliographic Services.

\bibliographystyle{iopart-num}
\bibliography{paper}

\end{document}